# A Novel Energy Stable Numerical Scheme for Navier-Stokes-Cahn-Hilliard Two-phase Flow Model with Variable Densities and Viscosities

Xiaoyu Feng[1][0000-0002-0050-1919]   Jisheng Kou[2][0000-0003-0986-5900]

Shuyu Sun*[3][0000-0002-3078-864X]

[1] National Engineering Laboratory for Pipeline Safety, Beijing Key Laboratory of Urban Oil and Gas Distribution Technology, China University of Petroleum, Beijing 102249, China
[2] School of Mathematics and Statistics, Hubei Engineering University, Xiaogan 432000, China
[3] Computational Transport Phenomena Laboratory, Division of Physical Science and Engineering, King Abdullah University of Science and Technology, Thuwal 23955-6900, Saudi Arabia
`shuyu.sun@kaust.edu.sa`

**Abstract.** A novel numerical scheme including time and spatial discretization is offered for coupled Cahn-Hilliard and Navier-Stokes governing equation system in this paper. Variable densities and viscosities are considered in the numerical scheme. By introducing an intermediate velocity in both Cahn-Hilliard equation and momentum equation, the scheme can keep discrete energy law. A decouple approach based on pressure stabilization is implemented to solve the Navier-Stokes part, while the stabilization or convex splitting method is adopted for the Cahn-Hilliard part. This novel scheme is totally decoupled, linear, unconditionally energy stable for incompressible two-phase flow diffuse interface model. Numerical results demonstrate the validation, accuracy, robustness and discrete energy law of the proposed scheme in this paper.

## 1   Introduction

Two-phase flow is omnipresent in many natural and industrial processes, especially for the petroleum industry, the two-phase flow is throughout the whole upstream production process including oil and gas recovery, transportation and refinery, e.g.[1].

As a critical component in the two-phase fluid system, the interface is usually considered as a free surface, its dynamics is determined by the usual Young-Laplace junction condition in classical interface tracking or reconstruction approaches, such as Level-set[2], volume-of-fluid[3] and even some advanced composite method like VOSET[4].

But when it traced back to 19th century, Van der Waals[5] provided a new alternative point of view that the interface has a diffuse feature, namely non-zero thickness. It can be implicitly characterized by scalar field, namely phase filed, taking constant values in the bulk phase areas and varying continuously but radically across a diffuse front. Within this thin tran-sition domain, the fluids are mixed and store certain quantities of "mixing energy" in this region. Thus, unlike other methods proposed graphically, phase dynamics is derived from interface physics by energy variational approach regardless of the numerical solution, which gives rise to coupled nonlinearly well-posed system at partial differential equation continuous form that satisfies thermodynamically consistent energy dissipation laws. Then there is possibility for us to design numerical scheme preserving energy law in discrete form[6].

The diffuse interface approach excels in some respects of handling two-phase flow among other available methods. Firstly, it is based on the principle of energy minimization. Hence it can deal with moving contact lines problems and morphological changes of interface in a natural way effortlessly with GNBC[7], such as droplet coalescence or break-up phenomena. Secondly, we can benefit from the simplicity of formulation, ease of numerical implementation without explicitly tracking or reconstructing interface, also the capability to explore essential physics at the interfacial domain. The accessibility of modeling various material properties or complex interface behaviors directly by introducing appropriate energy functions. Therefore, enforcing certain rheological fluid or modeling polymeric solutions or viscoelastic behaviors would be alluring feature naturally. For these benefits, the diffuse interface model attracted substantial academic attention in recent years, a great number of the advanced and cutting-edge researches are conducted corresponding to partial immiscible multi-components flow or flow with surfactant based on phase field theory[8,9] and thermodynamically consistent diffuse interface model for two-phase flow with thermo-capillary[10] et al.

The classical diffuse interface model for cases of two-phase incompressible viscous Newtonian fluids is known as the model H[11]. It has been successfully applied to simulate flows involving incompressible fluids with same densities for both phase components. This model is restricted to the matched density case using Boussinesq approximation. Unlike the matched density case, when it comes to the case with big density ratio, the incompressibility cannot gurantee mass conservation any longer in this model. Therefore, the corresponding diffuse interface model with the divergence free condition no longer preserve an energy law. Thus, a lot of further works have been done by (1998)Lowengrub[12], (2002)Boye[13], (2007)Ding[14], (2010)Shen [15] and most recently Benchmark computations were carried out by (2012)Aland[16].

Generally there are two kinds of approaches to deal with variable densities problem, one is that material derivative of momentum equation written in one kind form that takes density variations into consideration without resorting to mass conservation to guarantee stability of energy proposed by J.-L. Guermond [17]. Another approach is proposed by Abels[18]. The approach introduces an intermediate velocity to decouple the Cahn-Hilliard equation and Navier-Stokes equation system in Minjeaud's paper[19], and recently this approach is applied in Shen[20] to simulate the model in [18]. However, the schemes proposed in [19,20] employ the intermediate velocity in the Cahn-Hilliard equation only, imposing the mass balance equation to ensure the discrete energy-dissipation law. Very recently, in Kou[21] the schemes that the intermediate velocity is applied in both mass balance equations and the momentum balance equation are developed to simulate the multi-component diffuse-interface model proposed in [22] to guarantee the consistency between the mass balance and energy dissipation. In this paper, we extend this treatment to the model in [18]. However, this extension is not trivial due to a crucial problem that Cahn-Hilliard equation is not equivalent to mass balance equation. In order to deal with this problem, a novel scheme applying the intermediate velocity in Navier-Stokes equation will be proposed in this paper.

The rest part of this paper is organized as follows. In Section 2 we introduce a diffuse interface model for two-phase flow with variable densities and viscosities in detail; In Section 3 we propose a brand new numerical scheme for solving the coupled Navier–Stokes and Cahn–Hilliard equation system based on this model; In Section 4 some numerical results are demonstrated in this part to validate this scheme comparing with benchmark and to exam the accuracy, discrete energy decaying tendency. Other cases and numerical performances will be investigated to show the robustness of the novel scheme.

## 2 Mathematical formulation and physical model

The phase field diffuse interface model with variable densities and viscosities can be described through the following Cahn-Hilliard equation coupled with Navier-Stokes equation. An introduced phase field variable $\phi$, namely order parameter, defined over the domain, identifies the regions occupied by the two fluids.

$$\phi(\mathbf{x}, t) = \begin{cases} 1 & fluid\ 1 \\ -1 & fluid\ 2 \end{cases} \tag{1}$$

With a thin smooth transition front of thickness ε bridging two fluids, the microscopic interactions between two kinds of fluid molecules rules equilibrium profiles and configurations of interface mixing layer neighboring level-set $\Gamma_t = \{\phi: (\mathbf{x}, t) = 0\}$. For the situation of isotropic interactions, the following Ginzburg–Landau type of Helmholtz free energy functional is given by the classical self-consistent mean field theory in statistical physics[23]:

$$W(\phi, \nabla\phi) = \lambda \int_\Omega \left(\frac{1}{2}\|\nabla\phi\|^2 + F(\phi)\right) dx \tag{2}$$

The foremost term in right hand side represents the effect of mixing of interactions between the materials, and the latter one implies the trend of separation. Set the Ginzburg–Landau potential in the usual double-well form $F(\phi) = \frac{\phi^2 - 1}{4\varepsilon^2}$. $\lambda$ means mixing energy density, ε is the capillary width of interface between two phases. If we focus on one-dimensional interface and assume that total diffusive mixing energy in this domain equals to traditional surface tension coefficient:

$$\sigma = \lambda \int_{-\infty}^{+\infty} \left\{\frac{1}{2}\left(\frac{d\phi}{dx}\right)^2 + F(\phi)\right\} dx \tag{3}$$

The precondition that diffuse interface is at equilibrium is valid, then we can get the relationship among surface tension coefficient σ, capillary width ε and mixing energy density $\lambda$:

$$\sigma = \frac{(2\sqrt{2})}{3}\frac{\lambda}{\varepsilon} \tag{4}$$

The evolution of phase field ($\phi$) is governed by the following Cahn-Hilliard equations:

$$\begin{cases} \phi_t + \nabla \cdot (\mathbf{u}\phi) = M\Delta\mu \\ \mu = \frac{\delta W}{\delta \phi} = f(\phi) - \Delta\phi \end{cases} \qquad (5)$$

Where $\mu$ represents the chemical potential, namely $\frac{\delta W}{\delta \phi}$ that indicates the variation of the energy $W$ with respect to $\phi$; the parameter $M$ is a mobility constant related to the diffusivity of bulk phases and $f(\phi) = F'(\phi)$

The momentum equation for the two-phase system is presented as the usual form

$$\begin{cases} \rho(\mathbf{u}_t + (\mathbf{u} \cdot \nabla)\mathbf{u}) = \nabla \cdot \tau \\ \tau = \eta D(\mathbf{u}) - p\mathbf{I} + \boldsymbol{\tau}_e \\ \boldsymbol{\tau}_e = -\lambda(\nabla\phi \otimes \nabla\phi) \end{cases} \qquad (6)$$

with the identical equation

$$\nabla \cdot (\nabla \mathbf{u} \otimes \nabla \mathbf{u}) = (\Delta\phi - f(\phi)) + \frac{1}{2}\nabla(\|\nabla\phi\|^2 + F(\phi)) = -\mu\nabla\phi$$

$$+\frac{1}{2}\nabla(\|\nabla\phi\|^2 + F(\phi)) = \phi\nabla\mu + \frac{1}{2}\nabla(\|\nabla\phi\|^2 + F(\phi) - \phi\mu) \qquad (7)$$

The second term in eqn.(7) can be merged with the pressure gradient term $p$, then the pressure of the momentum equation should be a modified pressure $p^*$, It denotes that: $p^* = p + \frac{1}{2}\nabla(|\nabla\phi|^2 + F(\phi) - \phi\mu)$. The $p$ is represented the modified one in the following contents for unity.

## 2.1 Case of matched density

The governing equations system:

$$\begin{cases} \phi_t + \nabla \cdot (\mathbf{u}\phi) = M\Delta\mu \\ \mu = f(\phi) - \Delta\phi \\ \mathbf{u}_t + (\mathbf{u} \cdot \nabla)\mathbf{u} = \nabla \cdot \eta D(\mathbf{u}) - \nabla p - \phi\nabla\mu \\ \nabla \cdot \mathbf{u} \end{cases} \qquad (8)$$

A set of appropriate boundary condition and initial condition is applied to the above system: no-slip boundary condition for momentum equation and the period boundary condition for the Cahn-Hilliard equation.

$$\mathbf{u}|_{\partial\Omega} = 0 \quad \left.\frac{\partial\phi}{\partial n}\right|_{\partial\Omega} = 0 \quad \left.\frac{\partial\mu}{\partial n}\right|_{\partial\Omega} = 0 \qquad (9)$$

Also the initial condition:

$$\mathbf{u}|_{t=0} = \mathbf{u}_0 \quad \phi|_{t=0} = \phi_0 \qquad (10)$$

If the density contrast of two phases is relatively little, a common approach is to employ the Boussinesq approximation[24], replacing momentum equation in equation system(8) by

$$\rho_0(\mathbf{u}_t + (\mathbf{u} \cdot \nabla)\mathbf{u}) = \nabla \cdot \eta D(\mathbf{u}) - \nabla p - \phi\nabla\mu + \frac{\phi+1}{2}\delta\rho\mathbf{g} \qquad (11)$$

Set the background density $\rho_0 = (\rho_1 + \rho_2)$ and term $\frac{\phi+1}{2}\delta\rho\mathbf{g}$ is an additional body force term in charge of the equivalent gravitational effect caused by density difference. Since the density $\rho_0$ distributed everywhere in this field do not change respect to time. If the divergence of the velocity field $\nabla \cdot \mathbf{u}=0$ holds. Then basic mass conservation $\rho_t + \nabla \cdot (\rho\mathbf{u})=0$ is a natural consequence of incompressibility.

By inner product operation of 1st 2nd and 3rd equation of system (8) with $-\mu, \phi_t, \mathbf{u}$ respectively and summation of these three results, It is easily to conclude that system (8) admits the following energy dissipation law:

$$\frac{d}{dt}\int_\Omega \left(\frac{1}{2}\|\mathbf{u}\|^2 + \frac{\lambda}{2}\|\nabla\phi\|^2 + \lambda F(\phi)\right) dx = -\int_\Omega \left(\frac{\eta}{2}\|D(u)\|^2 + M\|\nabla\mu\|^2\right) dx \quad (12)$$

## 2.2 Case of variable density

Now we consider the case where the density ratio is so large that the Boussinesq approximation is no longer in effect. Here introduced a phase field diffuse interface model for incompressible two-phase flow with different densities and viscosity proposed by Abels[18].

$$\begin{cases} \phi_t + \nabla \cdot (\mathbf{u}\phi) = M\Delta\mu \\ \mu = f(\phi) - \Delta\phi \\ \rho\mathbf{u}_t + (\rho\mathbf{u} + \mathbf{J}) \cdot \nabla\mathbf{u} = \nabla \cdot \eta D(\mathbf{u}) - \nabla p - \phi\nabla\mu \\ \nabla \cdot \mathbf{u} = 0 \end{cases} \quad (13)$$

among them

$$\mathbf{J} = \frac{\rho_2 - \rho_1}{2} M\nabla\mu \quad (14)$$

The density and viscosity is the function of phase parameter.

$$\rho(\phi) = \frac{\rho_1 - \rho_2}{2}\phi + \frac{\rho_1 + \rho_2}{2} \quad \eta(\phi) = \frac{\eta_1 - \eta_2}{2}\phi + \frac{\eta_1 + \rho\eta_2}{2} \quad (15)$$

The mass conservation property can be derived from eqn. (14),(15) and (16).

$$\rho_t + \nabla \cdot (\rho\mathbf{u}) + \nabla \cdot \mathbf{J} = 0 \quad (16)$$

The NSCH governing system holds thermodynamically consistency and energy law. We can obtain the following energy dissipation law:

$$\frac{d}{dt}\int_\Omega \left(\frac{\rho}{2}\|\mathbf{u}\|^2 + \frac{\lambda}{2}\|\nabla\phi\|^2 + \lambda F(\phi)\right) dx = -\int_\Omega \left(\frac{\eta}{2}\|D(u)\|^2 + M\|\nabla\mu\|^2\right) dx \quad (17)$$

If we add the gravity in this domain, such as modeling topological evolution of a single bubble rising in a liquid column, the total energy must contain the potential energy. Then this energy dissipation law can be expressed as follow:

$$\frac{dE_{tot}}{dt} = \frac{d}{dt}\int_\Omega \left(\frac{\rho}{2}\|\mathbf{u}\|^2 + \frac{\lambda}{2}\|\nabla\phi\|^2 + \lambda F(\phi) - \rho\mathbf{g}y\right) dx \leq 0 \quad (18)$$

## 3 Decoupled Numerical scheme

### 3.1 Time discretization

In matched density case, for simplicity of presentation, we will assume that $\eta_1 = \eta_2 = \eta$. Given initial conditions $\phi^0, \mathbf{u}^0, p^0$ we compute $\phi^{k+1}, \mu^{k+1}, p^{k+1}, \widetilde{\mathbf{u}}^{k+1}, \mathbf{u}^{k+1}$ for n $\geq$ 0. Here is an additional term to the convective velocity introduced based on the idea from [19]. Then the intermediate velocity term $\hat{\mathbf{u}}^k = \mathbf{u}^k - \delta t \frac{\phi^k \nabla \mu^{k+1}}{\rho^k}$ makes the Cahn-Hilliard equation and the Navier-Stokes equation decoupled fundamentally. The novel scheme can be described as below:

$$\begin{cases} \frac{\phi^{k+1}-\phi^k}{\delta t} + \nabla \cdot (\hat{\mathbf{u}}^n \phi^{k+1}) = M\Delta\mu^{k+1} \\ \mu^{k+1} = \lambda(f_e(\phi^k) + f_c(\phi^{k+1}) - \Delta\phi^{k+1}). or. \\ \mu^{k+1} = \lambda(f(\phi^k) - \Delta\phi^{k+1}) + \frac{\lambda}{\varepsilon^2}(\phi^{k+1} - \phi^k) \\ \mathbf{n} \cdot \nabla\phi^{k+1}|_{\partial\Omega} = 0 \quad \mathbf{n} \cdot \nabla\mu^{k+1}|_{\partial\Omega} = 0 \end{cases} \quad (19)$$

We add a stabilizing term $\frac{\lambda}{\varepsilon^2}(\phi^{k+1} - \phi^k)$ or treat the term by convex splitting method $f(\phi) = f_e(\phi) + f_c(\phi)$. Then the time step for computation will not be strictly limited in extreme range by the coefficient Capillary width $\varepsilon$

$$\begin{cases} \rho_0 \frac{\widetilde{\mathbf{u}}^{k+1}-\mathbf{u}^k}{\delta t} + (\mathbf{u}^k \cdot \nabla)\widetilde{\mathbf{u}}^{k+1} = \nabla \cdot \eta_0 D(\widetilde{\mathbf{u}}^{k+1}) - \nabla p^k - \phi^k \nabla \mu^{k+1} + F_m \\ \widetilde{\mathbf{u}}^{k+1}|_{\partial\Omega} = 0 \end{cases} \quad (20)$$

Then we can get the pressure by solving a constant coefficient Poisson equation and correct the velocity filed to satisfy divergence free condition.

$$\begin{cases} \rho_0 \frac{\mathbf{u}^{k+1}-\widetilde{\mathbf{u}}^{k+1}}{\delta t} = -\nabla(p^{k+1} - p^k) \\ \nabla \cdot \mathbf{u}^{k+1} = 0 \\ \mathbf{u}^{k+1}|_{\partial\Omega} = 0 \end{cases} \quad (21)$$

For variable density case, [17] and [18] serve as incentive for the novel numerical scheme below. To deal with the variable densities and ensure numerical stability, we have to define a cut-off function $\hat{\phi}$ for the phase order parameter at first place.

$$\phi = \begin{cases} \phi & |\phi| \leq 1 \\ sign(\phi) & |\phi| > 1 \end{cases} \quad (22)$$

Given initial conditions $\phi^0, \mathbf{u}^0, p^0, \rho_0, \eta_0$ we compute $\phi^{k+1}, \mu^{k+1}, p^{k+1} \mathbf{u}^{k+1}, \rho^{k+1}, \eta^{k+1}$ for n $\geq$ 0. The discretization of the Cahn-Hilliard part is same with the matched density as equation (19)

We update the density and viscosity by cut-off function

$$\begin{cases} \rho(\phi^{n+1}) = \frac{\rho_1-\rho_2}{2}\hat{\phi}^{n+1} + \frac{\rho_1+\rho_2}{2} \\ \eta(\phi^{n+1}) = \frac{\eta_1-\eta_2}{2}\hat{\phi}^{n+1} + \frac{\eta_1+\rho_2}{2} \end{cases} \quad (23)$$

For the momentum equation part, we use

$$\begin{cases} \rho^k \frac{\mathbf{u}^{k+1}-\mathbf{u}^k}{\delta t} + (\rho^k \hat{\mathbf{u}}^k + \mathbf{J}^k) \cdot \nabla \mathbf{u}^{k+1} = \nabla \cdot \eta D(\mathbf{u}^{k+1}) - \nabla p^k \\ \qquad -\phi^k \nabla \mu^{k+1} + \frac{1}{2}(\frac{\rho_1+\rho_2}{2} \nabla \cdot \hat{\mathbf{u}}^k)\mathbf{u}^{k+1} \\ \qquad \mathbf{u}^{k+1}|_{\partial\Omega} = 0 \end{cases} \quad (24)$$

together with

$$\mathbf{J}^k = \frac{\rho_2-\rho_1}{2} M \nabla \mu^k \quad (25)$$

For saving the computer time consuming and stability, we adopt schemes based on pressure stabilization.

$$\begin{cases} \Delta(p^{k+1} - p^k) = \frac{\theta}{\delta t} \nabla \cdot \mathbf{u}^{n+1} \\ \nabla(p^{k+1} - p^k)|_{\partial\Omega} = 0 \\ \theta = \frac{1}{2}\min(\rho_2, \rho_1) \end{cases} \quad (26)$$

Pressure stabilization at the initial stage might cause velocity without physical meaning because it cannot satisfy solenoidal condition strictly. If we use pressure correction method, we have to face the non-linear Poisson equation. Thus, the solution must cost much more time.

### 3.2  Spatial discretization

For 2-D cases, the computational domain is $\Omega = (0, L_x) \times (0, L_y)$, the staggered grid are used for spatial discretization. The cell centers are located on

$$x_i = \left(i - \frac{1}{2}\right)h_x \quad i = 1, \ldots, n_x \qquad y_i = \left(j - \frac{1}{2}\right)h_y \quad i = 1, \ldots, n_y$$

Where $h_x$ and $h_y$ are grid spacing in $x$ and $y$ directions. $n_x$, $n_y$ are the number of grids along $x$ and $y$ coordinates respectively. In order to discretize the coupled Cahn-Hilliard and Navier-Stokes system, the following finite volume method is introduced.

$$U_h = \left(x_{i-\frac{1}{2}}, y_j\right) \Big| i = 1, 2, \ldots, n_x + 1; \quad j=1, 2, \ldots, n_y$$

$$V_h = \left(x_i, y_{j-\frac{1}{2}}\right) \Big| i = 1, 2, \ldots, n_x; \quad j=1, 2, \ldots, n_y + 1$$

$$P_h = \left(x_{i-\frac{1}{2}}, y_j\right) \Big| i = 1, 2, \ldots, n_x; \quad j=1, 2, \ldots, n_y$$

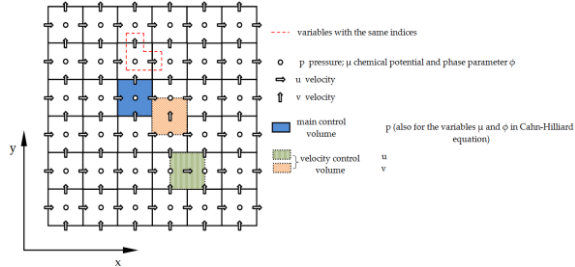

**Fig. 1.** The staggered grid based on finite volume method

Where $P_h$ is cell-centered space, $U_h$ and $V_h$ are edge-center space, $(\phi, \mu, p, \rho, \eta) \in P_h$, $u \in U_h$, $v \in V_h$. Some common differential and averaged operators are used to interpolation of these physical variables from one space to another space which are not discussed in detail here.

## 4     Numerical Results

### 4.1    Validation of the novel scheme

**Case1: a bubble rising in liquid in 2D domain**

There is a rectangular domain $\Omega = (0,1) \times (0,2)$ filled with two-phase incompressible fluid and a lighter bubble (with density $\rho_1$ and dynamic viscosity $\eta_1$) in a heavier medium (with density $\rho_2$ and dynamic viscosity $\eta_2$) rises from a fixed position initially. As it described in benchmark test paper[25], the boundary condition imposed on the vertical walls and top wall are replaced by free slip condition. The physical parameters for case 1 follows Table 1.

**Table 1.** The physical parameters for numerical test case 1.

| Case | $\rho_1$ | $\rho_2$ | $\eta_1$ | $\eta_2$ | g | σ | M |
|---|---|---|---|---|---|---|---|
| 1 | 1000 | 100 | 10 | 1 | 0.98 | 24.5 | $1\times 10^{-5}$ |

The initial bubble is perfect round with radius r = 0.25 and its center is set at the point $(x_0, y_0) = (0.5,1)$. The initial profile of ϕ is set as

$$\phi(x, y) = -\tanh(\frac{1}{\sqrt{2}\varepsilon}(\sqrt{(x - x_0)^2 + (y - y_0)^2} - r)) \quad (27)$$

We must note these parameters have been through the non-dimensionalization. Mobility coefficient is an additional numerical parameter, which is not appeared in the sharp interface model. The value is chosen in a rational range for comparison of different spatial step and interface thickness. Furthermore, the interface thick $\varepsilon$ is chosen proportional to $h$. The energy density parameter $\lambda$ can be calculated by the surface tension coefficient $\sigma$ through the equation (4). The time step $\delta t = 0.0001$.

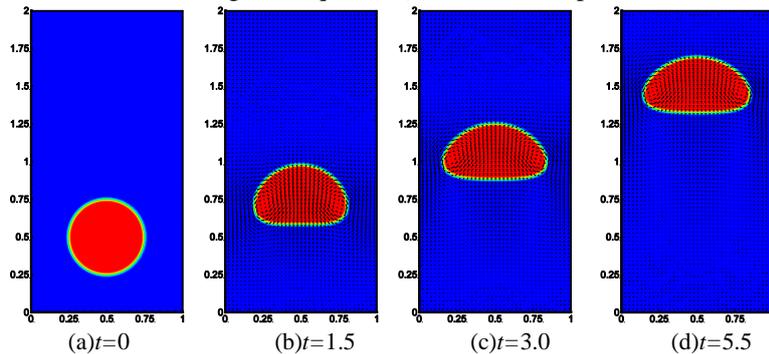

(a)*t*=0     (b)*t*=1.5     (c)*t*=3.0     (d)*t*=5.5

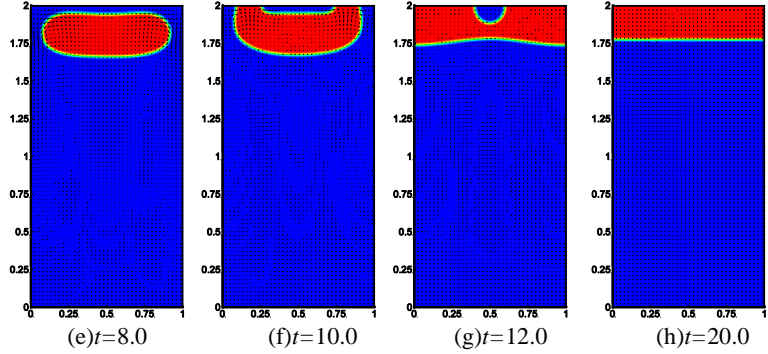

(e) $t=8.0$　　(f) $t=10.0$　　(g) $t=12.0$　　(h) $t=20.0$

**Fig. 2.** Snapshots of the bubble evolution with velocity vector field (computed on $h=1/150$ grid).

The bubble shape at the time point $t=3.0$ calculated by the novel scheme is compared with the solution from the benchmark paper by the level-set method[25] and the diffuse interface method[16] in Fig 3(a). Different bubble shapes at the time $t=3.0$ are compared from the coarsest grid ($h = 1/50, \varepsilon = 0.02$) to the finest grid ($h = 1/200, \varepsilon = 0.005$).

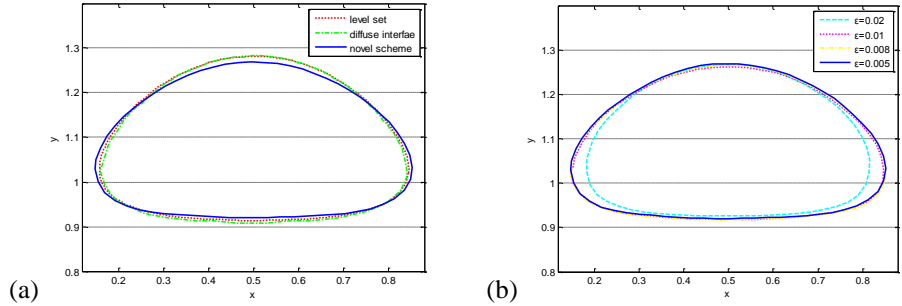

(a)　　(b)

**Fig. 3.** Bubble shapes at $t=3$ for the novel scheme comparing with the level-set and diffuse interface benchmark results provided in [16,25](a) ; bubble shapes at $t=3$ solved by grid with different refinement level and different interface thickness(b).

The shapes of bubble differ distinctly for different values of interface thickness $\varepsilon$. But they seem to be convergent so that there is no significant differences for the finest grid and the case with $\varepsilon = 0.008$. We can also remark that the bubble shape from novel scheme is quite approximate to the benchmark level-set and diffuse interface results. But it is clearly not sufficient to only look at the bubble shapes, therefore we use some previously defined benchmark quantities to validate the new scheme rigorously.

　　$I$. *Center of the mass:*

Various positions of points can be used to track the motion of bubbles. The most common way is to use the center of mass defined by

$$y_c = \frac{\int_{\phi>0} y dx}{\int_{\phi>0} 1 dx} \tag{28}$$

with $y$ as the vertical coordinate of $\mathbf{x}(x, y)$

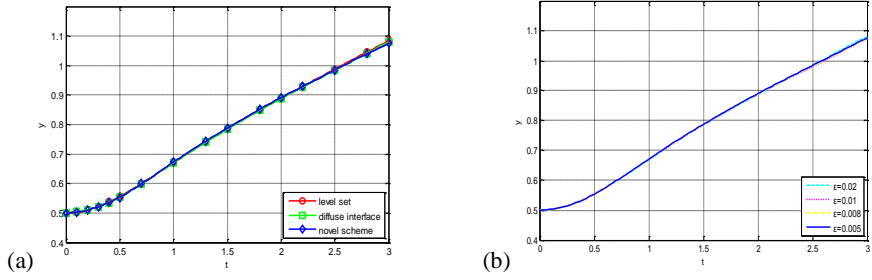

**Fig. 4.** Center of the mass of the bubble for the novel scheme comparing with the level-set and diffuse interface benchmark result provided in [16,25](a) ; Solutions of center of mass from grids with different refinement level and different interface thickness(b).

### II. Rise velocity

$v$ is the vertical component of the bubble's velocity $\mathbf{u}$. Where $\phi > 0$ denotes the region that bubble occupies. The velocity is volume average velocity of bubble.

$$v_c = \frac{\int_{\phi>0} v dx}{\int_{\phi>0} 1 dx} \tag{29}$$

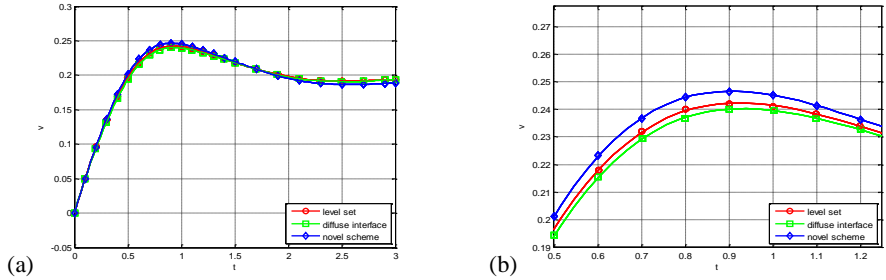

**Fig. 5.** Rise velocity of the bubble for the novel scheme comparing with the level-set and diffuse interface benchmark results (a); close-up of rise velocity at maximum values area (b).

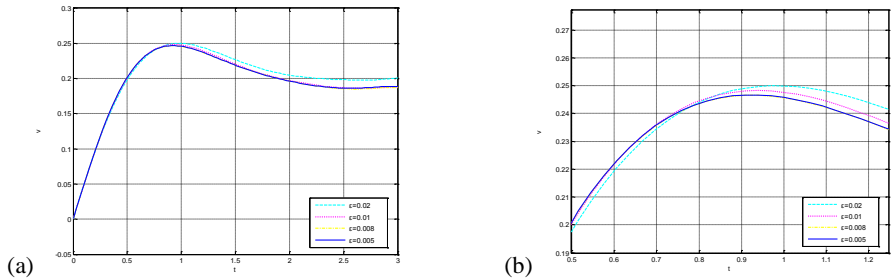

**Fig. 6.** Solutions of rise velocity of the bubble from grids with different refinement level and different interface thickness (a); close-up of rise velocity at maximum values area (b).

Combining the contours given in Fig 3 and Fig 4 , 5 ,6 about the mass center and rise velocity of the lighter bubble before $t$=3, a significant increase in rise velocity can be observed at the initial stage and then the velocity decrease slowly to a constant value when the time advances to $t$=3 gradually. The shape of bubble will reach to a temporary steady state when the rise velocity keeps constant. The mass center of the bubble could be recognized as a linear function of time asymptotically after it is higher than $y$=0.6.

Although the difference is visible for the coarsest grid and thickest interface on the velocity plot. It is obvious to see that results become closer and closer to the line corresponding to the computation on the finest grid with refinement. The results calculated by the new scheme shows good agreement with the level-set solution[25] and benchmark diffuse interface method[16].

The decaying trend of discrete energy in Fig 7 confirms that the proposed scheme is energy stable. The whole system reaches the equilibrium state at $t$ =25.

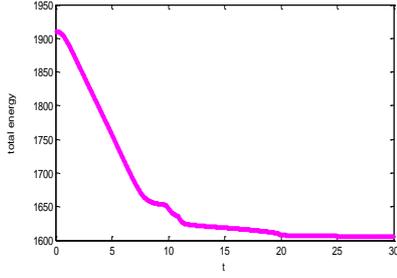 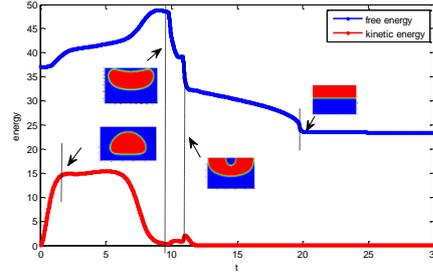

**Fig. 7.** Energy dissipation of the whole system   **Fig. 8.** Variation of free energy and kinetic energy of the whole system

Fig 8 gives the evolution of free energy and kinetic energy respectively. At the early stage of bubble rising, kinetic energy and free energy rise dramatically, which come from part of the reduced gravity potential energy. Then the velocity keep constant to some extent at next stage. The bubble shape also change into a relative steady state. When the bubble almost touching the top lid. The kinetic energy gives a considerable decrease to the zero. Then the gas phase will evolve to a stratified state finally under the lead of the diffusion in the Cahn Hilliard equation.

**Case2: novel scheme for matched density with Boussinesq approximation**

In the section, We simulate a physical problem with matched viscosities and a relative low density contrast($\rho_1 = 1$ ; $\rho_2 = 10$) which ensures the Boussinnesq approximation is applicable. We set 2d bubble diameter $d = 1.0$, g = 1.0, $\eta_1 = \eta_2 = 0.1$, $\lambda = 4 \times 10^{-3}$, Mobility=0.02 and $\varepsilon = 0.008$. The incompressible fluid in a rectangular domain $\Omega = (0,2) \times (0,4)$ with initially a lighter bubble center located on $(x_0, y_0) = (1,1.5)$. These dimensionless parameters are set according to the cases in [15].The mesh size is $200 \times 400$ and the boundary condition at the vertical wall is no-slip boundary condition.

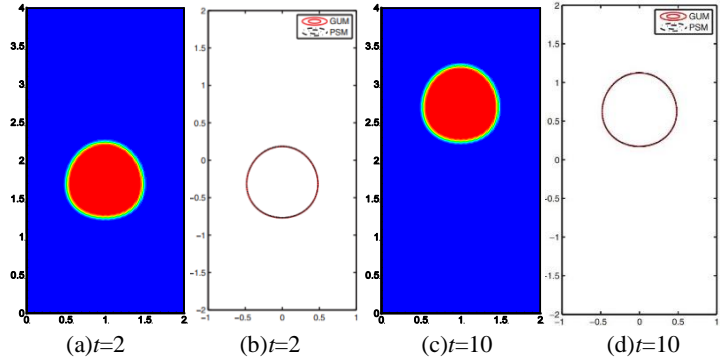

(a)*t*=2  (b)*t*=2  (c)*t*=10  (d)*t*=10

**Fig. 8.** The shape and displacement comparison of the bubble calculated by novel scheme with the results presented in [15] for matched density case.

It is easy to find that the shape of the bubble and the vertical displacement at different time step is pretty similar with the reference contour provided in [15] by GUM/PSM method. The reference only have a contour line at certain $\phi$. For a beautiful presentation of the integral contour calculated by the novel scheme here, we adjust interface thickness ε = 0.008. The novel scheme can be employed in the situation

### 4.2 Robustness test of the novel scheme

**Case3: Examining the performance for big density contrast**

We now consider two-phase incompressible fluid with the same initial condition and boundary condition with the case1. But density and viscosity contrast is much more violent in this case.

**Table 2.** The physical parameters for numerical test case 3.

| Case | $\rho_1$ | $\rho_2$ | $\eta_1$ | $\eta_2$ | g | σ | M |
|---|---|---|---|---|---|---|---|
| 3 | 1000 | 1 | 10 | 0.1 | 0.98 | 1.96 | $1\times10^{-5}$ |

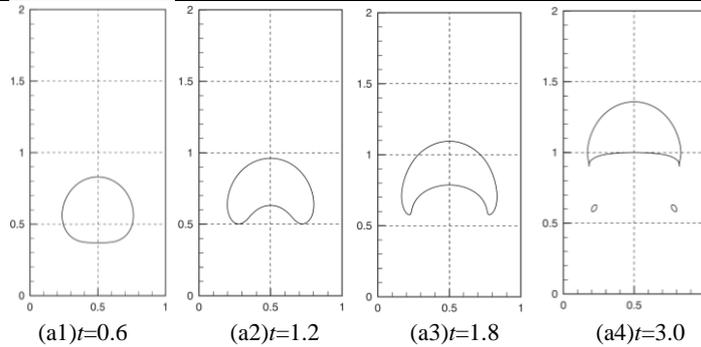

(a1)*t*=0.6  (a2)*t*=1.2  (a3)*t*=1.8  (a4)*t*=3.0

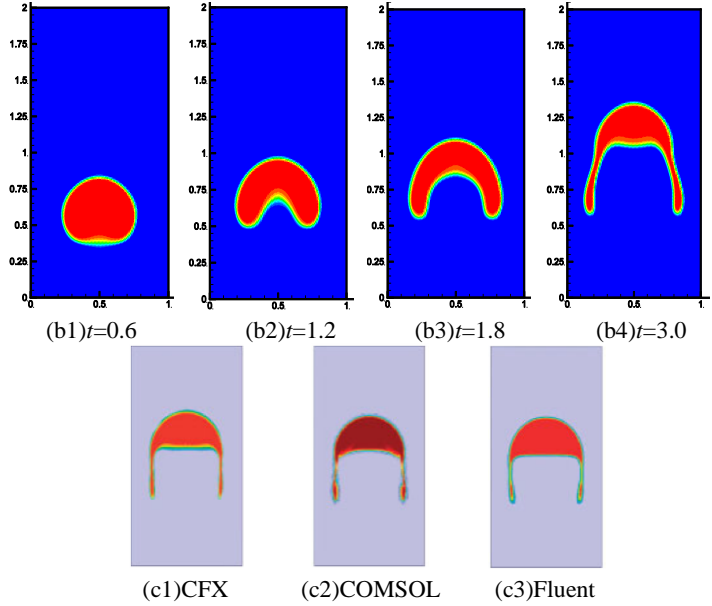

(b1)$t$=0.6   (b2)$t$=1.2   (b3)$t$=1.8   (b4)$t$=3.0

(c1)CFX   (c2)COMSOL   (c3)Fluent

**Fig. 8.** The shapes and displacement comparison of the bubble calculated by novel scheme with the benchmark level-set results and the contours at $t$=3.0 provided by three common commercial software in [25]

The break-up of bubble happen before the time $t$=3.0 in the level-set benchmark. So it is not appropriate to compare the results from the new scheme with the contour. But from the results of some common commercial computing software[25]. It's not that difficult to find the shape of bubble and the vertical displacement at $t$=3.0 solved by our scheme is pretty close to them. Although it could be some slight diffusion on the interface caused by the Cahn-Hilliard system itself. The case shows the robustness of the novel scheme proposed in this paper. It can not only handle an extreme numerical situation with harsh density and viscosity ratio but get reliable results to some extent.

## 5   Concluding remark

The numerical simulation and approximation of incompressible and immiscible two-phase flows with matched and variable densities and viscosities is the main topic in this paper. We proposed a brand new scheme for coupled diffuse interface model system with matched and variable densities and viscosities that satisfies the mass conservation and admits an energy law. Plenty of numerical experiments are carried out to illustrate the validation, accuracy compared with sharp interface method by the benchmark problem and to test the robustness of the new scheme for some extreme cases as well.

**References**

1. X.Y, Zhang., B, Yu.,et al.: Numerical study on the commissioning charge-up process of horizontal pipeline with entrapped air pockets. Advances in Mech Eng ,1-13 (2014).
2. M, Sussman., S, Osher.: A level-set approach for computing solutions to incompressible two-phase flow. Journal of Computational Physics 114,146–159 (1994).
3. C.W, Hirt., B.D, Nichols.: Volume of fluid (VOF) method for the dynamics of free boundaries. J. Computational. Physics 39,201–225(1981).
4. D.L, Sun., W.Q, Tao.: A coupled volume-of-fluid and leve-set (VOSET) method for computing incompressible two-phase flows. International Journal of Heat and Mass Transfer 53, 645–655(2010).
5. J.D, van der Waals., V, Konink.: The thermodynamic theory of capillarity under the hypothesis of a continuous density variation. J. Stat. Phys, Dutch (1893).
6. D.J, Eyre.: Unconditionally gradient stable time marching the Cahn-Hilliard equation. Computational and mathematical models of microstructural evolution, Mater. Res. Soc. Symp. Proc5. 529, 39-46 (1998).
7. G.P, Zhu, H.X, Chen, S.Y, Sun et al,. A fully discrete energy stable scheme for a phase filed moving contact line model with variable densities and viscosities. arXiv preprint arXiv:1801.08739, 2018.
8. G.P, Zhu, J.S, Kou J, S.Y, Sun et al. Numerical approximation of a binary fluid-surfactant phase field model of two-phase incompressible flow. arXiv preprint arXiv:1804.06305, 2018.
9. J.S, Kou., S.Y, Sun.: Multi-scale diffuse interface modeling of multi-component two-phase flow with partial miscibility. Journal of Computational Physics (1)318, 349–372 (2016).
10. Z, Guo., P, Lin.: A thermodynamically consistent phase-field model for two-phase flows with thermo-capillary effects. Journal of Fluid Mechanics 766, 226-271(2015).
11. M.E, Gurtin., D, Polignone., J, Vinals.: Two-phase binary fluids and immiscible fluids described by an order parameter, Math. Models Meth. Appl. Sci. 6,815–831(1996).
12. J, Lowengrub., L, Truskinovsky.: Quasi-incompressible Cahn–Hilliard fluids, Proc. R. Soc. London Ser. A 454 2617–2654(1998).
13. F. Boyer.: A theoretical and numerical model for the study of incompressible mixture flows, Comp. Fluids 31, 41–68(2002).
14. H, Ding ., C, Shu.: Diffuse interface model for incompressible two-phase flows with large density ratios, Journal of Computational Physics 226(2), 2078-2095(2007).
15. J, Shen ., X.F, Yang.: A phase-field model and its numerical approximation for two-phase incompressible flows with different densities and viscosities, SIAM J. Sci. Comput (3)32, 1159–1179(2010).
16. S, Aland ., A,Voigt.: Benchmark computations of diffuse interface models for two-dimensional bubble dynamics, Int. J. Numer. Meth. Fluids 69, 747–761(2012).
17. J.-L, Guermond ., L, Quartapelle.: A Projection FEM for Variable Density Incompressible Flows, Journal of Computational Physics 165,167–188 (2000).
18. H, Abels., H, Garcke and G, Grün.: Thermodynamically consistent, frame indifferent diffuse interface models for incompressible two-phase flows with different densities Math. Mod. Meth. Appl. Sic. 22(3), 1150013-1-40 (2012).
19. S, Minjeaud.: An Unconditionally Stable Uncoupled Scheme for a Triphasic Cahn- Hilliard/Navier-Stokes Model. Numerical Methods for Partial Differential Equations(2)29,584–618(2013).

20. J, Shen., X, Yang.: Decoupled, energy stable schemes for phase-field models of two-phase incom- pressible flows. SIAM Journal on Numerical Analysis (1)53,279-296(2015).
21. J, Kou., S, Sun., X, Wang.: Linearly decoupled energy-stable numerical methods for multi-component two-phase compressible flow, arXiv:1712.02222, (2017).
22. J, Kou., S, Sun.: Thermodynamically consistent modeling and simulation of multi-component two-phase flow with partial miscibility. Computer Methods in Applied Mechanics and Engineering 331, 623–649(2018).
23. P.M, Chaikin.,T.C, Lubensky.: Principles of Condensed Matter Physics, Cambridge University Press. United Kingdom(1995).
24. Boussinesq, Joseph.: Théorie de l'écoulement tourbillonnant et tumultueux des liquides dans les lits rectilignes a grande section. 1, Gauthier-Villars (1897).
25. Hysing, S., Turek, S, et al.: Quantitative benchmark computations of two-dimensional bubble dynamics, Int. J. Numer. Meth. Fluids 60, 1259–1288(2009).